\documentclass[%
 reprint,
onecolumn,
 amsmath,amssymb,
 aps,
prfluids,
]{revtex4-2}

\usepackage{graphicx}
\usepackage{dcolumn}
\usepackage{bm}

\usepackage{graphicx, amsmath, amssymb, amsfonts, mathtools, mathrsfs, color}
\usepackage{comment, todonotes}
\usepackage{soul}
\usepackage{dcolumn}

\usepackage{xcolor}
\usepackage[capitalise]{cleveref}
\usepackage[normalem]{ulem}
\usepackage{bm}

\usepackage{url}

\newcommand{\pd}[2]{ \frac{ \partial #1}{ \partial #2 } }

\newcommand{\pdi}[2]{ \partial_{#2} #1 }

\newcommand{\bvec}[1]{\ensuremath{\boldsymbol{#1}}}
\newcommand{\grad}{\nabla}

\newcommand{\Pra}{\text{Pr}}
\newcommand{\Ra}{\text{Ra}}
\newcommand{\Reyy}{\text{Re}}
\newcommand{\Nu}{\text{Nu}}
\newcommand{\RePr}{\Reyy{}\cdot\Pra{}}

\newcommand{\si}{\text{Supplemental Material}}

\newcommand{\uu}{\bvec{u}}

\newcommand{\Lap}{\grad^2}

\newcommand{\Area}{\text{A}_0}

\newcommand{\len}{h}

\newcommand{\xc}{X}
\newcommand{\yc}{Y}

\newcommand{\lrms}{L_{\rm rms}}
\newcommand{\Or}{\mathcal{O}}

\newcommand{\tavg}[1]{\langle #1 \rangle}

\begin{document}

\preprint{APS/123-QED}

\title{Low-order reaction-diffusion system approximates heat transfer and flow structure in annular convection}

\author{Yuejia Zhang}
 \affiliation{NYU-ECNU Institute of Physics and Institute of Mathematical Sciences, New York University Shanghai, Shanghai, 200124, China}
\author{Nicholas J. Moore}%
\email{nickmoore83@gmail.com}
\affiliation{Department of Mathematics, Colgate University, Hamilton, NY 13346, USA}%

\author{Jinzi Mac Huang}%
\email{machuang@nyu.edu}
 \affiliation{NYU-ECNU Institute of Physics and Institute of Mathematical Sciences, New York University Shanghai, Shanghai, 200124, China}
\affiliation{Applied Math Lab, Courant Institute, New York University, New York, NY 10012, USA}

\date{\today}

\begin{abstract}
Heat transfer in a fluid can be greatly enhanced by natural convection, giving rise to the nuanced relationship between the Nusselt number and Rayleigh number that has been a focus of modern fluid dynamics. Our work explores convection in an annular domain, where the  geometry reinforces the large-scale circulatory flow pattern that is characteristic of natural convection. The flow must match the no-slip condition at the boundary, leading to a thin boundary layer where both the flow velocity and the temperature vary rapidly.  To understand the system's heat transfer characteristics, we derive a reduced model from the Navier-Stokes-Boussinesq equations, whereby the equations of flow and heat are transformed to a system of low-order partial differential equations (PDEs) that take the form of a reaction-diffusion system. Solutions to the reaction-diffusion system, though they fail to predict dynamic events, preserve the same boundary-layer structure seen in the direct numerical simulation (DNS). By matching the solutions inside and outside the boundary layer, asymptotic analysis predicts a power-law relationship $\Nu{} \propto \Ra{}^{1/4}$.
Though difficult to distinguish from an exponent of $2/7$, the predicted power law agrees well with measurements from DNS over several decades of the Rayleigh number. Considering the model’s deficiencies in describing turbulent fluctuations and reversal events, the agreement regarding heat transfer characteristics is encouraging and suggests that the methodology of systematically deriving low-order PDEs from the governing equations may provide a useful complement to existing theories.
\end{abstract}

\maketitle


\section{Introduction}

Buoyancy variations resulting from the uneven heating of a fluid create a complex motion known as thermal convection. On a planetary scale, thermal convection brings motion to an otherwise motionless world, leading to atmospheric and oceanic flows \citep{Salmon1998, Zhong2009}, mantle and liquid-core convection \citep{Whitehead1972, zhang2000periodic, Zhong2005, Whitehead2015, mac2018stochastic}, solar magneto-hydrodynamics \citep{Wit2020}, and more \cite{mccurdy2022predicting, whiteheadenergy}.

As a relative measure between the buoyancy and viscous forcing, the Rayleigh number $\Ra{}$  dictates the dynamics of thermal convection: Below a critical number $\Ra{}<\Ra{}_1^*$, viscosity damps motion and the fluid conducts heat essentially as a solid would; At high $\Ra{}$, convective motion becomes turbulent, leading to dynamical features such as a large-scale circulation (LSC) and enhanced heat transfer \citep{RevModPhys.81.503}.

Dimensionlessly, heat transfer is characterized by the Nusselt number, which is the ratio of convective to conductive heat transfer. For $\Ra{}\leq\Ra{}_1^*$, heat transfer is largely conductive, giving $\Nu{} \approx 1$. For $\Ra{}\gg\Ra{}_1^*$, vigorous convective motion enhances the heat transfer considerably, and a power-law relationship $\Nu{}\propto \Ra^{\beta}$ has been observed in the range $\Ra{}=10^6$--$10^{14}$ \citep{niemela2000turbulent, funfschilling2005heat}.
In particular, the exponent $\beta = 2/7$ has been found to robustly describe laboratory and direct numerical measurements for $\Ra{}$ up to $10^{10}$ \cite{Belmonte1994,johnston2009comparison,Huang2022a,vieweg2023large}. This exponent has been rationalized theoretically by the mixing-zone model \citep{Castaing_1989}.
A theory developed by Grossmann and Lohse \citep{grossmann2000scaling, RevModPhys.81.503, grossmann_lohse_2013} incorporates the heat transfer contributed by both the thermal boundary layers (BL) and the bulk mixing to combine different scaling laws under a unifying framework. In the theory, different regimes of BL or bulk dominated thermal or kinetic dissipation give rise to different exponents. The exponent $2/7$ is found to be a good approximation to the linear combination of the regimes $\beta=1/4$ and $\beta=1/3$ with empirically estimated coefficients.
It has been conjectured that higher Rayleigh number gives rise to another regime in which the the heat flux in the thermal boundary layer is independent of domain height, giving $\beta = 1/3$.
This scaling law has seen support from large-scale direct numerical simulations up to $\Ra{} = 10^{15}$, albeit in a low aspect ratio domain \citep{iyer2020classical}. 
The above studies consider Rayleigh-Bénard convection (RBC) in which the upper and lower thermally driven surfaces are planar, e.g.~a rectangular domain or a cylindrical domain with the axis of rotation aligned with gravity.

Meanwhile, a large-scale circulation (LSC) can spontaneously develop in the limit of high $\Ra{}$, whereby the upwelling and downwelling motions combine to form a circulatory flow with overall direction determined by multiple factors such as domain geometry, external forces (e.g.~Coriolis), random perturbations, and more. The emergence and reversals of the LSC have been observed in controlled laboratory experiments \citep{Creveling1975, Gorman1984, Gorman1986, Castaing_1989, sreenivasan2002mean, Brown2007, Xi2007, Sugiyama2010, Song2011, Wang2018, Chen2019} and in numerical simulations \citep{Sugiyama2010, chandra2011dynamics, gallet2012reversals, chandra2013flow, Xu2021}, leading to many theories exploring its origin and control \citep{Lorenz1963,Araujo2005, Brown2007, chandra2013flow, Ni2015}. In particular, much attention has been devoted to the LSC developed in cubic and cylindrical domains: In the former case, 
the LSC can change its orientation through a reversal of the circulation direction \citep{Araujo2005}; For the latter case, a slow migration of the LSC orientation is possible \citep{Brown2005}. The possibility of influencing and controlling the LSC is also explored by modifying the thermal and flow boundary conditions \citep{zhang2021stabilizing,Huang2022a,mac2024side,zhang2025eliminating,PhysRevResearch.3.013231,ettel2025effectsconjugateheattransfer}, adding rotation \citep{PhysRevLett.102.044502,PhysRevLett.103.024503,ding2023vortex}, and inserting obstacles \citep{bao2015enhanced,li2024enhanced}. In different limits of domain aspect ratios, the LSC formation can either be more coherent and energetic \citep{PhysRevLett.111.104501,doi:10.1073/pnas.2403699121}, or be seen as a particular manifestation of the large-scale self-organization of the flow \citep{KAUFER2023283,Alam_2025}.

\begin{figure}
 \includegraphics[width = 0.8\textwidth]{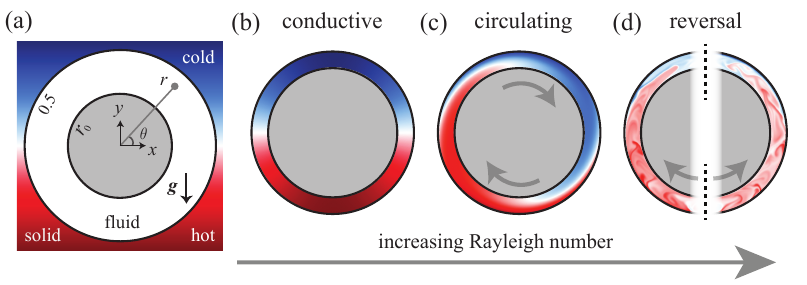}
 \centering
 \caption{Thermal convection in an annular domain. (a) The 2-dimensional annular fluid domain $r \in [r_0, \frac{1}{2}]$ is embedded in a solid background with constant vertical temperature gradient. (b) At low $\Ra{}$, the fluid is motionless and the heat only conducts. (c) Increasing $\Ra{}$ beyond critical, the fluid circulates unidirectionally. (d) Further increasing $\Ra{}$ leads to the reversals of LSC. In (b)-(d), $\Pra{} = 4$ and $r_0 = 0.4$, and the Rayleigh numbers are (b) $\Ra{}=1.4\times 10^5$, (c) $\Ra{}=2.2\times 10^6$, and (d) $\Ra{}=1.6\times 10^9$. Movies of (b)-(d) are included as \si{}.
 }
\label{fig1}
\end{figure}

Moore \& Huang (2024) \cite{Moore_Huang_2024} examined LSC dynamics in the annular domain illustrated in \cref{fig1}(a). 
The annulus offers some advantages in that the geometry itself reinforces the dominant circulatory nature of flow while suppressing other features, such as corner rolls or plumes that would tend to rise through the domain center.
The system is driven by a linear temperature profile imposed along the outer ring, which could be achieved by embedding the fluid in a  solid that is heated from below and that possesses a much higher conductivity (e.g.~water embedded in aluminum). As the Rayleigh number increases in this system, a sequence of dynamical states emerge, including a state of pure conduction [\cref{fig1}(b)], a state of steady circulation in either the clockwise or counterclockwise direction [\cref{fig1}(c)], and a state in which the LSC reverses direction spontaneously [\cref{fig1}(d)]. Numerical simulations of these states are included in \si{}. 
By exploiting simplifications made possible by the annular geometry, Moore \& Huang (2024) \cite{Moore_Huang_2024} systematically derived a low-dimensional ODE model from the governing Navier-Stokes Boussinesq equations. This model successfully recovers the range of dynamical states, including LSC reversals, with quantitative accuracy.

While the ODE model successfully predicts coarse-grained flow features and LSC reversals, it fails to accurately predict gross heat transfer rates due to its under-resolution of small-scale structures such as the thermal boundary layers. It is straightforward, however, to extend the model to a PDE system that fully resolves details in the radial direction, and, thus, resolves the thermal and momentum boundary layers that exist along the outer boundary of the annulus. The purpose of the present paper is to explore the capability of this extended model in predicting heat transfer rates.

Interestingly, heat transfer in annular convection may differ from the classical case of Rayleigh-Bénard convection (RBC), since the inner boundary of the annulus acts as an obstacle that impedes motion through the center. In this way, the annular geometry permits one to examine the interaction between the LSC and boundary layers in a controlled setting, without other confounding effects, such as thermal plumes rising through the center of the domain. To examine heat transfer in this geometry, we reduce the Navier-Stokes-Boussinesq equations to a set of low-dimensional PDEs whose solutions recover the boundary-layer structure of the full equations. The low-dimensional PDE takes the form of a reaction-diffusion system and predicts the Nusselt number to scale as $\Ra^{1/4}$. Comparison against DNS shows that this prediction, while difficult to distinguish from the closely related $\Ra^{2/7}$ scaling, does reasonably approximate heat transfer rates over a large range of $\Ra$. Furthermore, long-time solutions to the reaction-diffusion system are steady, suggesting that the majority of the $\Nu$--$\Ra$ relationship in an annulus can be understood purely as a boundary-layer phenomenon and the remaining discrepancy between the exponents $1/4$ and $2/7=0.286$ may be due to dynamic effects, such as LSC reversals, that are present in the DNS but not the reaction-diffusion system.

In what follows, we will outline the equations and DNS method in Sec. \ref{EQS}, and summarize the previous ODE model in Sec. \ref{ODE}. Next, the PDE model of annular convection is introduced in Sec. \ref{PDE}, whose boundary layer analysis is included in Sec. \ref{BoundaryLayer}. Finally, we will show how this PDE model recovers various scalings in Sec. \ref{Results}, and further discuss these results in Sec. \ref{discussion}.

\section{Equations and DNS method}
\label{EQS}

The dimensionless Navier-Stokes-Boussinesq equations for velocity $\uu$, pressure $p$, and temperature $T$ are
\begin{align}
\label{NS}
& \pd{\uu}{t} + \uu \cdot \grad \uu = -\grad p +
\Pra \Lap \uu + \Pra \, \Ra \, T \bvec{e_y}, \\
\label{trans}
& \pd{T}{t} + \uu \cdot \grad T = \Lap T, \\
\label{incomp}
& \grad \cdot \uu = 0.
\end{align}
Here, we have rescaled length by the domain height $\len$, time by the diffusive time scale $h^2/\kappa$ ($\kappa$ is the thermal diffusivity), and temperature by the maximum difference $\Delta  T$ between the top and bottom temperatures imposed on the outer ring. After rescaling, the annular domain of fluid is bounded between radius $r_0$ and $r_1 = 1/2$ as shown in \cref{fig1}(a), and the temperature lies in the range $T\in[0,1]$. Three dimensionless numbers arise:  the Rayleigh number $\Ra{} = (\beta_T \Delta T \len^3 g) / (\nu \kappa)$, the Prandtl number $\Pra = \nu/\kappa$, and the domain aspect ratio $r_0 = R_0/h$, where $\beta_T$ is the thermal expansion coefficient, $g$ is the acceleration due to gravity, $\nu$ is the kinematic viscosity, and $R_0$ is the dimensional radius of the inner boundary. 

In polar coordinates, $\uu = u\bvec{e}_\theta+ v \bvec{e}_r$ where $\bvec{e}_r$ and $\bvec{e}_\theta$ are the unit vectors in the $r$ and $\theta$ directions, the boundary conditions become 
\begin{align}
\label{noslip}
&u = v = 0 \hspace{25pt} \text{at } r=r_0 \text{ and } r=1/2, \\
\label{Tinner}
&\pd{T}{r} = 0 \hspace{35pt} \text{at } r=r_0, \\
\label{Touter}
&T = \frac{1-\sin \theta}{2} \hspace{10pt} \text{at } r=1/2 .
\end{align}

We use a pseudo-spectral  Chebyshev-Fourier method with implicit-explicit time stepping to solve \cref{NS,trans,incomp} with boundary conditions \cref{noslip,Tinner,Touter}. In particular, the method recasts \cref{NS,trans,incomp} in streamfunction-vorticity form \citep{peyret2002spectral, mac2021stable, Huang2022a}. The details of our numerical implementation can be found in \cite{Moore_Huang_2024}, where second-order convergence in time and spectral accuracy in space were verified. For the DNS conducted in this study, we set $r_0=0.4$ and $\Pra{}=0.5, 4, 16$. For each value of $\Pra{}$, we take $30$ different values of $\Ra{}$ so $\Ra{} = 10^8 \cdot 2^{k/2}$, where $k = -19, -18, \cdots, 10$. Thus $\Ra{} \in [1.4\times 10^5, 3.2\times 10^9]$, covering a wide range of dynamical behaviors shown in \cref{fig1}.

Without fluid motion ($\uu = \mathbf{0}$), it is easy to solve \cref{trans} and obtain the temperature distribution of the conductive state, 
\begin{equation}
    \label{tcond}
   T_{\text{cond}} = \frac{1}{2} -  \frac{1}{r}\left(\frac{r^2+r_0^2}{1+4r_0^2}\right)\sin{\theta}.
\end{equation}
We note that this $T_{\text{cond}}$ does not balance \cref{NS} and a flow must present near the inner boundary $r = r_0$ due to the Neumann boundary condition \cref{Tinner}. This flow is weak, leading to an $\Or{(1)}$ Reynolds number that is small compared to the flow strength of thermal convection. As this flow has negligible effects on the temperature distribution, we regard $T_{\text{cond}}$ as the conductive state in this study.

With fluid motion, we can define two dimensionless numbers: the Nusselt number measuring the heat transfer rate and the Reynolds number reflecting the flow rate, 
\begin{equation}
\label{NuRe}
    \Nu  = \frac{\langle \int_0^\pi (\partial_r T)|_{r=1/2}\, d\theta\rangle}{\int_0^\pi (\partial_r T_{\text{cond}})|_{r=1/2}\, d\theta}, \quad
    \Reyy =  \Pra^{-1} \langle \max |\uu|\rangle,
\end{equation}
where $\langle \cdot \rangle$ is the long-time average operation. For the Nusselt number, both the convective heat transfer rate (numerator) and the conductive heat transfer rate (denominator) are defined as the rate of heat flowing through the upper half boundary where $r = 1/2$ and $\theta \in (0,\pi)$. 

We introduce three coarse-grained variables to describe the heat and flow structures that emerge during annular convection, namely the fluid center of mass (CoM) $(\xc, \yc)$ and the fluid angular momentum $L$. The angular momentum describes, to leading order, the overall circulatory motion that arises. Meanwhile, $(\xc,\yc)$ quantifies how buoyancy variations alter the CoM. For example, heating the annulus from below, with no flow response, raises the CoM above the annulus center, $\yc>0$, whereas circulatory motion may offset the CoM horizontally. These variables are defined by,
\begin{equation}
\label{COM}
\xc = - \frac{1}{\Area} \int_{\Omega} r \, T \cos\theta\, dA \, , \quad
\yc = - \frac{1}{\Area} \int_{\Omega} r \,  T \sin\theta \, dA \, ,
\quad
L = \frac{1}{\Area}\int_{\Omega} r u \, dA ,
\end{equation}
where $\Area = \pi(1-4r_0^2)/4$ is the area of the annulus $\Omega$ and $dA = r\,dr d\theta$ is the area element.

\section{An ODE model recovering the LSC dynamics}\label{ODE}

Moore \& Huang (2024) derived an ODE system from the Navier-Stokes-Bousinesq equations that describes the evolution of the coarse-grained variables $\xc$, $\yc$, and $L$ \cite{Moore_Huang_2024}. Here, we briefly sketch the derivation of the ODE system in a way that will generalize to the new PDE model.

To begin, we expand each of the temperature $T$ and the flow velocity $\uu = u\bvec{e}_\theta+ v \bvec{e}_r$ fields as a Fourier series in $\theta$, 
\begin{align}\label{TFourier}
T(r,\theta,t) &= a_0(r,t) + \sum_{n=1}^{\infty}  a_n(r,t) \cos n\theta + b_n(r,t) \sin n\theta , \\
\label{uvFourier}
u(r,\theta,t) = &\sum_{n=-\infty}^{\infty} u_n(r,t) e^{i n \theta},  \quad
v(r,\theta,t) = \sum_{n=-\infty}^{\infty} v_n(r,t) e^{i n \theta}.
\end{align}

In the thin channel limit $r_0\to 1/2$, the dominant balance of \cref{incomp,noslip} implies $v\to0$ and $\pdi{}{\theta}u \to 0$, giving $u = u_0(r,t)$ and $v = 0$ at leading order. Making these substitutions and integrating the $u$ component of Navier-Stokes \cref{NS} over $\theta$ gives 
\begin{equation}
\label{u0dot}
    r^2 \pd{u_0}{t} = \frac{r^2}{2} \Pra{}\cdot\Ra{} \cdot a_1 + \Pra{}\left[ - u_0  + r \pd{}{r} \left( r \pd{u_0}{r}  \right)\right].
\end{equation}
Meanwhile, inserting the truncation $u = u_0(r,t)$ into \cref{trans} decouples the temperature modes, 
\begin{align}
\label{andot}
& r^2 \dot{a}_n = -n r \, u_0(r,t) \, b_n - n^2 a_n + r \pdi{}{r} \left( r \pdi{}{r} a_n \right) , \\
\label{bndot}
& r^2 \dot{b}_n = +n r \, u_0(r,t) \, a_n - n^2 b_n + r \pdi{}{r} \left( r \pdi{}{r} b_n \right) .
\end{align}
This equation holds for $n = 0, 1, 2, \cdots$.

The boundary conditions for \cref{u0dot,andot,bndot} are 
\begin{align}
    \label{abu-cond-0}
    &u_0 = \pdi{}{r} a_n = \pdi{}{r} b_n = 0  &&\mbox{at } r = r_0,\\
    \label{abu-cond-1}
    &u_0 = 0,\, a_0 = 1/2,\, b_1 = -1/2,\, \mbox{all others vanish} &&\mbox{at } r = 1/2.
\end{align}
for $n = 0, 1, 2, \cdots$.

By representing $u_0$, $a_1$, and $b_1$ each as a truncated Laurent series in $r$ and inserting into \cref{u0dot,andot,bndot}, Moore \& Huang (2024) obtained the following dynamical system for the evolution of $(L, \xc, \yc)$ \cite{Moore_Huang_2024}:
\begin{align}
\label{Ldot2}
&\dot{L} = - \Ra\, \Pra \, \xc - \alpha \Pra \, L , \\
\label{xdot2}
&\dot{\xc} = -k L (\yc - y_1)  - \beta \xc , \\
\label{ydot2}
&\dot{\yc} = +k L \xc - \beta (\yc - y_0) ,
\end{align}
where $\alpha, \beta, k, y_0, y_1$ are non-negative coefficients that depend on  geometry alone (i.e.~independent of $\Ra{}$ and $\Pra{}$).

\begin{figure}
 \includegraphics[width = 0.75\textwidth]{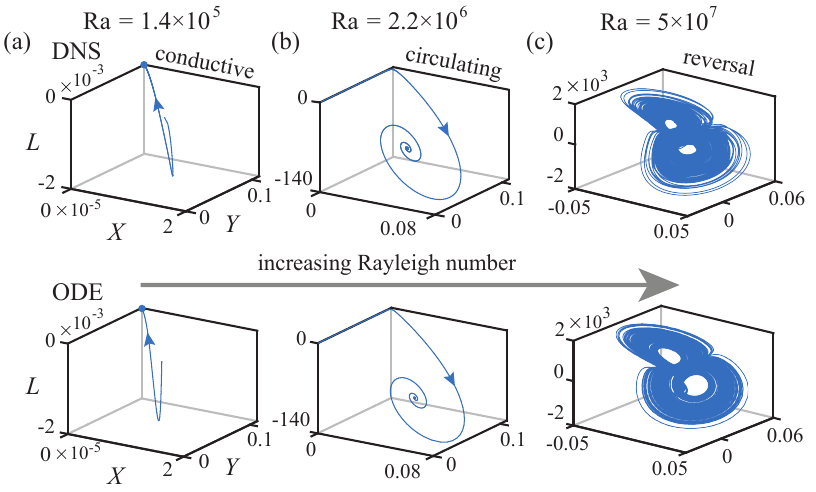}
 \centering
 \caption{Trajectories of $(L, \xc, \yc)$ obtained from the DNS result and the ODE model. (a) Conductive state at $\Ra{}=1.4\times 10^5$. (b) Circulating state at $\Ra{}=2.2\times 10^6$. (c) Reversal state at $\Ra{}=5\times 10^7$. In all simulations, $\Pra{}=4$ and $r_0=0.4$.}
\label{fig2}
\end{figure}

Linear stability analysis of \cref{Ldot2,xdot2,ydot2} reveals two critical Rayleigh numbers: $\Ra_1^*$ marks the loss of stability of the conductive state and the simultaneous emergence of the (bistable) circulating states, while $\Ra_2^*$ marks the loss of stability of the circulating state and the subsequent onset of LSC reversals \citep{Moore_Huang_2024}. Exact formulas for $\Ra_1^*$ and $\Ra_2^*$ are provided in \cite{Moore_Huang_2024}. To briefly demonstrate the transitions, \cref{fig2} shows trajectories of $(L(t), X(t), Y(t))$ computed from the DNS (top) and from the ODE system (bottom) for three Rayleigh numbers. In this figure, the parameters $\Pra{}=4$ and $r_0=0.4$ are fixed, yielding $\Ra^*_1{}=7.3\times 10^5$ and $\Ra^*_2{}= 1.6\times 10^7$. \Cref{fig2}(a) shows the case $\Ra{} < \Ra^*_1{}$, giving convergence to the stable conductive state in both DNS and the ODE model. \Cref{fig2}(b) features the intermediate case, $\Ra^*_1{}<\Ra{} < \Ra^*_2{}$, resulting in convergence to steady circulation. \Cref{fig2}(c) shows the high Rayleigh number case, $\Ra^*_2{}<\Ra{}$, which gives chaotic LSC reversals in both the DNS and ODE system. More detailed discussion of these transitions can be found in \citep{Moore_Huang_2024}.

\begin{figure}
 \includegraphics[width = 0.9\textwidth]{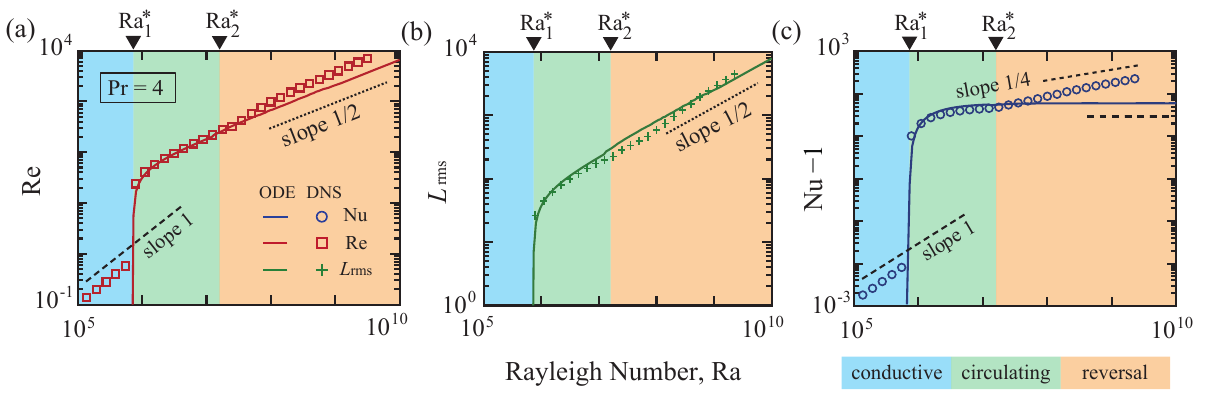}
 \centering
 \caption{ $\Reyy{}$, $\lrms{}$, $\Nu{}$ measured from the DNS and the ODE model. Reynolds number (a) and fluid angular momentum (b) are small when $\Ra{}<\Ra{}_1^*$ but have a 0.5 power-law scaling at high $\Ra{}$. Both the DNS (symbols) and the ODE (curves) solutions capture this scaling. (c) Nusselt number is slightly above unity for the conductive state $\Ra{} < \Ra{}_1^*$ and has a $1/4$ power-law scaling for high $\Ra{}$. In this limit, the ODE model approaches a constant $\Nu{}$, thus it fails to capture the heat transfer of annular convection. In all simulations, $\Pra{}=4$ and $r_0=0.4$, and two critical Rayleigh numbers are identified as $\Ra{}^*_1 = 7.3\times 10^5$ and $\Ra{}^*_2 = 1.6\times 10^7$. }
\label{fig3}
\end{figure}

While \cref{fig2} shows that the ODE model captures the trajectories of $(L(t), X(t), Y(t))$ with surprising detail, we now evaluate other features of the model, including those that quantify the heat transport. We focus on the coarse-grained variables $\Nu{}$ and $\Reyy{}$, defined in \cref{NuRe}. In the DNS,  these quantities are computed directly from \cref{NuRe}. We also compute the root-mean-square average of angular momentum, $\lrms{} = \sqrt{\langle L^2 \rangle}$, which reflects the overall strength of the LSC. For the ODE model, insertion of the truncated fields $u=u_0(r,t)$ and $T = 1/2+ a_1(r,t) \cos\theta + b_1(r,t) \sin\theta$ into \cref{NuRe} yields the formulas
\begin{align}
\label{NuRe_ODE} 
\Reyy{} = &\frac{12}{\Pra{}}\left(\frac{1-\sqrt{2r_0}}{1+\sqrt{2r_0}}\right)\lrms{}, \\
\Nu{} = &\frac{48(1+2r_0)(1+4r^2_0)}{(1-2r_0)^2(1+6r_0+16r^2_0)}\left[\frac{1+2r_0+4r_0^2}{12(1+2r_0)}- \tavg{Y} \right].
\label{Nu_formula}
\end{align}
where $\tavg{Y}$ is the mean height of the fluid CoM. Physically speaking, vigorous convection both intensifies thermal transport and lowers the fluid CoM by allowing warm fluid to rise more easily. As such, $\tavg{Y}$ and $\Nu{}$ are negatively related, as borne out by \cref{Nu_formula}.

\Cref{fig3} shows the three coarse-grained quantities, $\Reyy{}$, $\lrms{}$, and $\Nu{}$, plotted against $\Ra$ for both the DNS (symbols) and the ODE model (curves) with $\Pr=4$. The Reynolds number and LSC strength in \cref{fig3}(a)-(b) show good agreement between the ODE and the DNS: For $\Ra{} > \Ra{}_1^*$, a common power law with exponent 0.5 exists as \cref{NuRe_ODE} indicates that $\Reyy{}$ and $\lrms{}$ are proportional to each other. The scaling $\Reyy{} \sim \Ra{}^{0.5}$ is also present in the classic RBC \citep{RevModPhys.81.503}, and we will later verify this scaling through a dominant-balance analysis. However, deviation appears in the conductive regime $\Ra < \Ra_1^*$, where the DNS has a small but non-vanishing $\Reyy{}$ while the ODE model is motionless. This is due to the fact that the conductive temperature distribution \cref{tcond} yields a weak flow in the Navier-Stokes-Boussinesq \cref{NS}, whose strength is proportional to $\Ra{}$ due to the buoyancy term there. Indeed, \cref{fig3}(a) shows that $\Reyy{} = \Or{(\Ra)}$ for $\Ra < \Ra_1^*$, and rapidly jumps to a much higher value when $\Ra{}$ crosses $\Ra{}_1^*$. The ODE model,  on the other hand, has only one stable equilibrium that is associated with $\Reyy{} = 0$.  

\Cref{fig3}(c) shows the heat transfer rate as quantified by the Nusselt number, $\Nu{}$. Below the threshold $\Ra{} < \Ra{}_1^*$, $\Nu{}$ is identically one for the ODE model but slightly above unity for the DNS -- once again due to the fluid motion associated with the conductive state. For $\Ra{} > \Ra{}_1^*$, $\Nu{}$ increases with $\Ra{}$ as convective motion sets in and then grows in strength. The ODE model predicts the $\Nu{}$-$\Ra{}$ behavior reasonably well in the range $\Ra{}_1^* < \Ra{} < \Ra{}_2^*$, but for $\Ra{} > \Ra{}_2^*$ there is considerable discrepancy. The DNS shows that $\Nu{}$ continues to grow as $\Ra{}$ increases, with the measured data points suggesting the power law $\Nu{}\sim \Ra{}^{1/4}$. The ODE model, however, predicts $\Nu{}$ to saturate to a constant value at large $\Ra{}$. In fact, \cite{Moore_Huang_2024} shows that the value of $\tavg{Y}$ predicted by the ODE-model becomes independent of $\Ra$ if $\Ra$ is sufficiently large, which through \cref{Nu_formula} implies the horizontal asymptote of $\Nu$ seen in \cref{fig3}(c). This discrepancy in the behavior of $\Nu$  suggests that the ODE model may have oversimplified the temperature field by neglecting fine-scale structure in thermal gradients. To address this deficiency, we next present a PDE model that accurately resolves the thermal boundary layer structure of annular convection.

\section{Reaction-diffusion PDE model}
\label{PDE}

Rather than expanding the $r$-dependence of the main state variables in a Laurent series, we observe that truncating \cref{u0dot,andot,bndot} at the first Fourier mode in $\theta$ and retaining the full dependence on $r$ produces a closed PDE system,
\begin{align}    
\label{pde-u}
\pd{u_0}{t} &= \frac{1}{2} \Pra{} \, \Ra{} \, a_1 + \frac{\Pra{}}{r^2}\left[ - u_0  + r \pd{}{r} \left( r \pd{u_0}{r}  \right)\right].\\
\label{pde-a}
\pd{a_1}{t} &= -\frac{u_0 b_1 }{r} + \frac{1}{r^2}\left[- a_1 + r \pd{}{r} \left( r \pd{a_1}{r}\right)\right] , \\
\label{pde-b}
\pd{b_1}{t} &= +\frac{u_0 a_1 }{r} + \frac{1}{r^2}\left[-  b_1 + r \pd{}{r} \left( r \pd{b_1}{r}  \right) \right] ,
\end{align}
\Cref{abu-cond-0,abu-cond-1} give boundary conditions,
\begin{align}
\label{bd0}
    &u_0 =0,\quad  \partial_r a_1 =0,\quad  \partial_r b_1 = 0 &&\mbox{at } r = r_0,\\
    \label{bd1}
    &u_0 = 0,\quad  a_1 = 0,\quad b_1 = -1/2 &&\mbox{at } r = 1/2.
\end{align}

Interestingly, \cref{pde-u,pde-a,pde-b} take the form of a {\em reaction-diffusion} (RD) system, with the nonlinear reaction terms $-u_0 b_1/r$ and $u_0 a_1/r$ representing convective transport of heat. The temperature field feeds back on momentum as seen in the term $\Pra{} \, \Ra{} \, a_1 /2$ in \cref{pde-u}, and all three components diffuse as seen in the right-most terms of each equation.
This system can be solved by standard numerical methods \citep{peyret2002spectral,trefethen2000spectral}. In particular, we use a pseudo-spectral Chebyshev-Fourier method to discretize $u_0(r,t)$, $a_1(r,t)$, $b_1(r,t)$ in space and an implicit-explicit method to step forward in time \citep{Moore_Huang_2024}.

\begin{figure}
 \includegraphics[width = 0.75\textwidth]{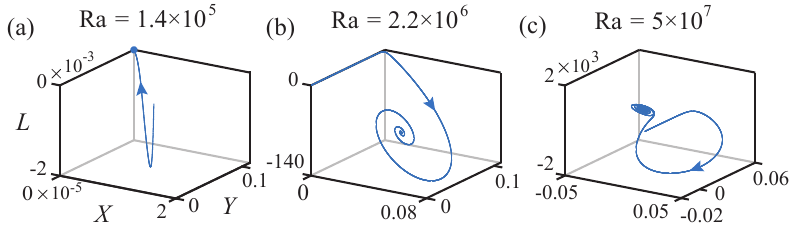}
 \centering
 \caption{Trajectories of $(L, \xc, \yc)$ obtained from the RD model, \cref{pde-u,pde-a,pde-b,bd0,bd1}. (a) Conductive state at $\Ra{}=1.4\times 10^5$. (b) Circulating state at $\Ra{}=2.2\times 10^6$. (c) At $\Ra{}=5\times 10^7$, the RD solution converges to a circulating state instead of a chaotic reversal state. In all simulations, $\Pra{}=4$ and $r_0=0.4$.}
\label{fig4}
\end{figure}

From the numerical solution of \cref{pde-u,pde-a,pde-b,bd0,bd1}, we can compute the quantities $L(t)$, $\xc(t)$, and $\yc(t)$ via \cref{COM}. \Cref{fig4} shows the trajectories of ($L$, $\xc$, $\yc$) at the same three values of $\Ra{}$ used in \cref{fig2}. \Cref{fig4}(a)-(b) shows that the RD model accurately recovers the dynamics in the conductive (a) and the circulating (b) regimes, agreeing with both the ODE model and the full DNS. We see in \cref{fig4}(c), however, that the RD model fails to predict chaotic LSC reversals; solutions exhibit a single reversal and then convergence to a circulating state. It appears the damping effect of diffusion in \cref{pde-u,pde-a,pde-b} overcomes what would otherwise be chaotic dynamics. This situation results from the choice to fully resolve the $r$-dependency while leaving $\theta$-dependency truncated to the lowest order in \cref{pde-u,pde-a,pde-b}. Despite this weakness in predicting coarse-grained {\em dynamics}, we will see that the new RD model considerably improves the prediction of {\em time-averaged thermal transport} as quantified by $\Nu$.

Seeing as the numerical solutions of \cref{pde-u,pde-a,pde-b,bd0,bd1} always exhibit convergence to a steady state, regardless of the $\Ra$ value, we hereafter focus on the steady-state quantities $a_1 = a_1(r)$, $b_1 = b_1(r)$, and $u_0 = u_0(r)$. \Cref{fig5} provides a visualization of these steady-state solutions at three different Rayleigh numbers. The bottom row shows the profiles $a_1(r)$, $b_1(r)$, and $u_0(r)$ directly, while the top row shows the corresponding temperature and velocity fields computed via $T = 1/2 + a_1(r) \cos{\theta} + b_1(r) \sin{\theta}$ and $u = u_0(r)$. In the first case of $\Ra{} < \Ra{}_1^*$ [\cref{fig5}(a) and (d)], the conductive state $a_1 = 0, b_1 = -(r + r_0^2 r^{-1})/(1 + 4r_0^2), u_0 = 0$ corresponding to \cref{tcond} is stable and there is no fluid motion in the RD solution. Increasing $\Ra$ beyond $\Ra_1^{*}$ gives rise to the circulating state with nontrivial flow field, as first seen in \cref{fig5}(b) and (e). At even higher Rayleigh number, Figs.~\ref{fig5}(c) and (f) show that a circulating state still emerges as a steady state, but, interestingly, the profiles show a distinct boundary-layer structure. That is, sharp temperature and flow variations appear only in a narrow region surrounding the outer boundary $r=1/2$, where the thermal forcing is imposed. As seen in the next section, this emergent boundary-layer structure leads to accurate predictions of the thermal transport, despite the absence of reversal dynamics in the RD model.

\section{Boundary Layer Analysis of the Reaction-Diffusion Model}
\label{BoundaryLayer}

\begin{figure}
 \includegraphics[width = 0.7\textwidth]{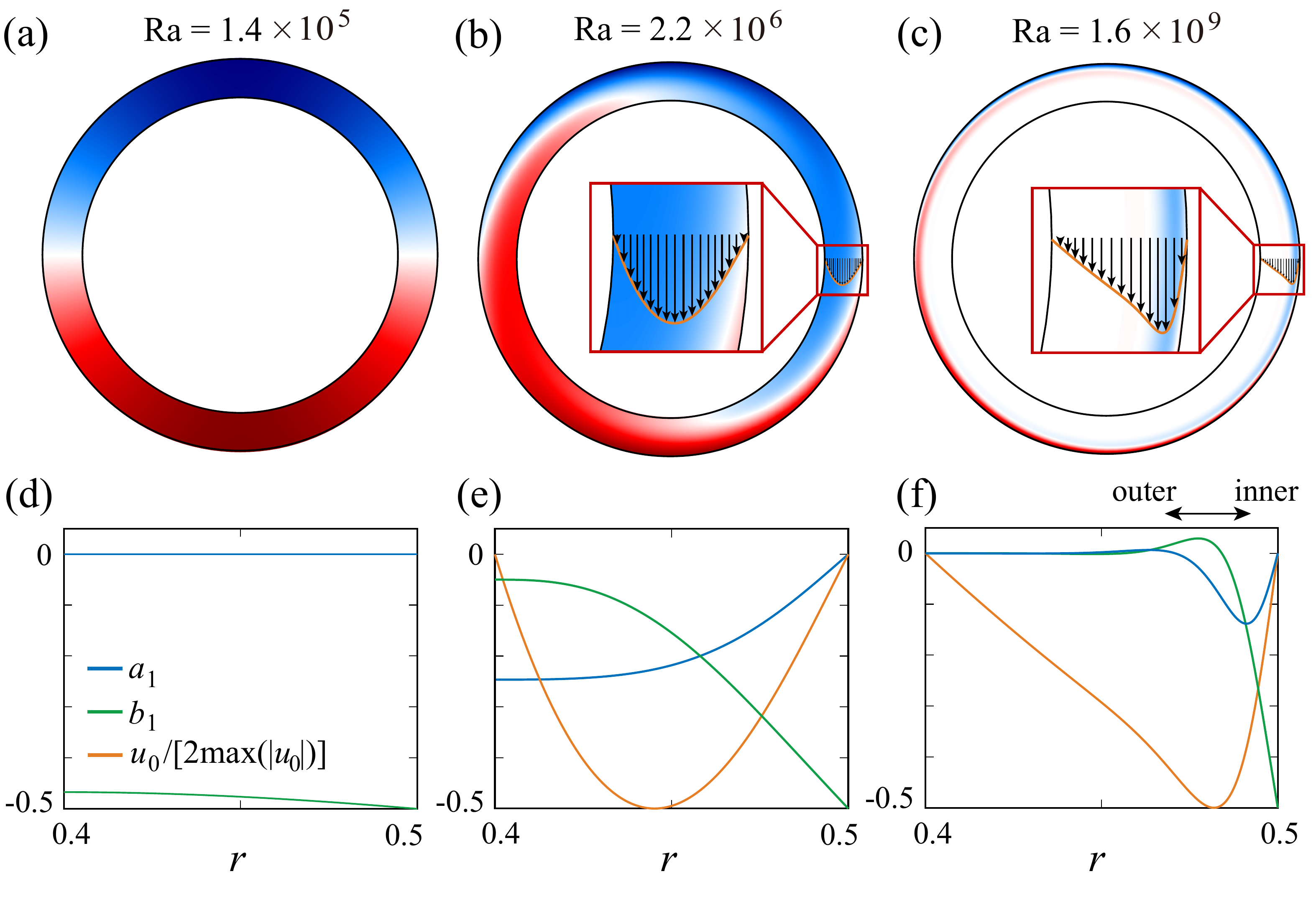}
 \centering
 \caption{Steady-state RD model solutions of the flow and temperature fields. (a)-(c) show the RD model solutions with increasing $\Ra{}$. Thermal and momentum boundary layers develop at high $\Ra{}$ as shown in the zoom-in view (inset) of (c). (d)-(f) show the distribution of $a_1$, $b_1$, and $u_0$ corresponding to (a)-(c). In all simulations, $\Pra{} = 4$ and $r_0 = 0.4$.}
\label{fig5}
\end{figure}

With the aim of predicting thermal transport, we now conduct boundary-layer analysis of the reaction-diffusion PDE system, \cref{pde-u,pde-a,pde-b,bd0,bd1}, in the limit of high $\Ra$. The characteristic boundary layer thickness $\delta \ll 1$, marks the so-called inner region near $r = 1/2$ where sharp temperature and flow variations dominate; see \cref{fig5}(c) and (f).
We first analyze this region by introducing the stretched coordinate $R = (1/2-r)/\delta$. Rewriting \cref{pde-u,pde-a,pde-b} in terms of variables $A(R) = a_1(r(R))$, $B(R) = b_1(r(R))$, and $U(R) = u_0(r(R))$ gives
\begin{align}
    \delta^{-2}A'' + \frac{2\delta^{-1}}{1-2R\delta}A' + \frac{4}{(1-2R\delta)^2}A &= \frac{2}{1-2R\delta} U B,\label{innereq_a_full}\\ \delta^{-2}B'' + \frac{2\delta^{-1}}{1-2R\delta}B' + \frac{4}{(1-2R\delta)^2}B &= - \frac{2}{1-2R\delta}  U A, \label{innereq_b_full}\\\delta^{-2}U'' + \frac{2\delta^{-1}}{1-2R\delta}U' + \frac{4}{(1-2R\delta)^2}U &= -\frac{1}{2}\Ra{}A, \label{innereq_u_full}
\end{align}
Since $\delta \ll 1$, the first term on each left-hand side (with prefactor $\delta^{-2}$) dominates the second and third terms. The only possible non-trivial balance as $\delta \to 0$ is therefore
\begin{align}
    \delta^{-2}A'' = 2\ UB,\quad 
    \delta^{-2}B''  = - 2\ UA,\quad \delta^{-2}U'' = -\frac{1}{2}\ \Ra{}\ A.\label{innereq}
\end{align}
Meanwhile, \cref{bd0,bd1} imply boundary conditions 
\begin{align}
\label{innerbd_inner}
U(0) &= A(0) = 0, \, B(0) = -1/2 .
\end{align}
Far-field ($R \to \infty$) conditions on $U, A, B$ would be obtained by matching to the outer solution.

The nonlinear system of ODEs given by \cref{innereq} does not appear to be amenable to exact solution and so we consider asymptotic behavior for $\Ra \gg 1$. Assuming  scaling relationships $A = \Or(\Ra{}^{\mu})$, $B = \Or(\Ra{}^{\nu})$, $U = \Or(\Ra{}^{\lambda})$, $\delta = \Or(\Ra{}^{\epsilon})$ and inserting into \cref{innereq} yields the three algebraic equations,
\begin{align}
\label{alg_eqs}
\mu = \nu, \quad 
\lambda = -2\epsilon, \quad
-4\epsilon = 1+ \mu.
\end{align}
Since there are four unknown exponents, this system is under-determined.

A fourth condition arises from knowledge of the steady-state Fourier coefficients $A$ and $B$. Due to \cref{trans,incomp}, the steady-state temperature field $\lim_{t \to \infty} T(r,\theta,t)$ satisfies a maximum principle. In particular, it is limited by the values from \cref{Touter} imposed on the Dirichlet boundary, $0 \le \lim_{t \to \infty} T(r,\theta, t) \le 1$
We remark that this bound does not hold for finite time because the initial temperature distribution may have values outside of the range $[0,1]$. 
Nonetheless, the bound on the steady-state temperature field implies that the magnitude of the Fourier coefficients $A$ and $B$ are bounded above, independent of $\Ra$. Therefore, these variables are at most $\Or(1)$ with respect to $\Ra \gg 1$, giving $\mu, \nu \le 0$. Further, the condition $B(0) = -1/2$ from \cref{innerbd_inner} implies that $\nu = 0$. With this extra piece of information, the unique solution to \cref{alg_eqs} is $\nu=0$, $\mu=0$, $\lambda = 1/2$, $\epsilon = -1/4$, which gives the following scaling relations for $\Ra \gg 1$,
\begin{equation}
\label{scaling}
A = \Or(1),\, B = \Or(1),\, U = \Or(\Ra{}^{1/2}),\, \delta = \Or(\Ra{}^{-1/4}) .
\end{equation}
In particular, the scaling of the boundary layer thickness $\delta = \Or(\Ra{}^{-1/4})$, along with $A, B = \Or(1)$, gives an estimate of the boundary heat flux, $\partial_r T |_{r=1/2} = \Or(\delta^{-1}) = \Or(\Ra^{1/4})$, which, through definition \cref{NuRe}, implies the following scaling law for the Nusselt number,
\begin{equation}
\label{Nu-scaling}
\Nu{}\propto \Ra{}^{1/4} \quad \mbox{ for } \Ra \gg 1.
\end{equation}

We next consider the outer region in which $r$ is sufficiently separated from the boundary $r=1/2$. In this region, the flow velocity $u_0$ must match $U$ as $r\to 1/2$, giving $u_0 = \Or(\Ra^{1/2})$.
Recall that $a_1, b_1$ are bounded in magnitude by the maximum principle. 
Inserting the scaling $u_0 = \Or{(\Ra^{1/2})}$ into \cref{pde-a,pde-b}, the variables $a_1$ and $b_1$ must be (strictly) asymptotically smaller than $\Or(\Ra^{-1/2})$ in order to avoid a contradiction; that is, $a_1, b_1 = o(\Ra^{-1/2})$. Inserting this information into \cref{pde-u} gives a variable-coefficient ODE satisfied by $u_0$,
\begin{align}
r \pd{}{r} \left( r \pd{u_0}{r}  \right) - u_0 =0
\end{align}
The solution that satisfies the correct matching condition is $u_0 = \Ra^{1/2} C (r-r^2_0/r)$ where $C$ is an $\Or(1)$ constant that would in principle be determined by the matching procedure. 
\Cref{fig5}(f) provides numerical confirmation of the asymptotic shape of the flow profile $u_0$ as well as the fact that $a_1, b_1 \to 0$ as $\Ra \to \infty$.

An immediate consequence of this analysis is that the characteristic flow speed $U$ increases as the square root of the Rayleigh number, as is consistent with the classic RBC results \citep{RevModPhys.81.503}. Furthermore, inserting the scaling of $U$ into the definitions \cref{NuRe} gives the scaling laws
\begin{align}
\label{ReyL-scaling}
\Reyy{} \propto \Pra{}^{-1} \Ra{}^{1/2}, \quad
\lrms{} \propto \Ra{}^{1/2},     
\end{align}
which, along with \cref{Nu-scaling}, can be tested against DNS.

\section{Results}
\label{Results}

\begin{figure}
 \includegraphics[width = \textwidth]{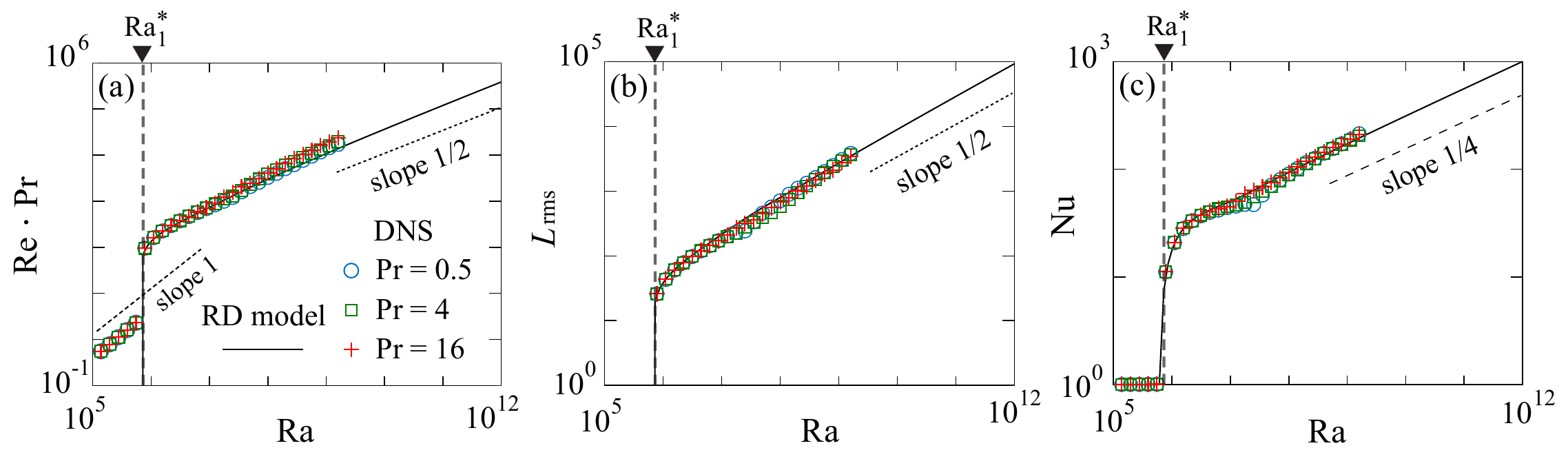}
 \centering
 \caption{$\Reyy{}$, $\lrms{}$, and $\Nu{}$ obtained from the DNS (symbols) and RD model (curves) solutions. (a) The Reynolds number has a scaling $\Reyy{}\propto \Pra{}^{-1} \Ra{}^{1/2}$. (b) Fluid angular momentum scales as $\lrms{}\propto \Ra{}^{1/2}$. (c) Nusselt number follows a power-law with an exponent around 1/4. Here $r_0 = 0.4$, leading to $\Ra{}^*_1 = 7.3\times 10^5$.}
\label{fig6}
\end{figure}

We now aim to test the predictions offered by the RD model, \cref{pde-a,pde-b,pde-u}, especially scaling laws \cref{Nu-scaling,ReyL-scaling} that result from boundary-layer analysis. As previously discussed, the RD model does not accurately describe temporally sensitive features, such as LSC reversal events. Nonetheless, it may offer improved predictions for time-averaged quantities, such as thermal transport, due to its ability to resolve small boundary layers in $r$. Accordingly, \cref{fig6} shows measurements of three time-averaged quantities: $\RePr$, $\lrms{}$, and $\Nu{}$. The figure shows these values as measured in DNS of the full Navier-Stokes-Boussinesq equations (symbols), as well as those computed from numerically solving the RD system (solid curves). As seen in the figure, results from DNS and from the RD model agree closely with one another for $\Ra>\Ra^*_1$, with only slight discrepancies between the two. In particular, \cref{fig6}(a) shows that the product $\RePr$ collapses all of the data from different $\Pra{}$ simulations onto a single master curve, which is indeed well predicted by the RD model. \Cref{fig6}(b) confirms that the RD model also recovers the mean angular momentum accurately. Further, the scaling law \cref{ReyL-scaling}, predicts both $\RePr$ and $\Pra{}$ to scale as $\Ra^{1/2}$, as confirmed by comparison with the dotted line. 

For $\Ra<\Ra^*_1$, the RD model predicts a motionless conductive state while the DNS reveals that there is actually a weak flow present. This flow is due to the misalignment between gravity and isopycnals that intersect the inner insulating boundary.
As a result, the RD data in \cref{fig6}(a) show $\Reyy{} = 0$ for $\Ra<\Ra^*_1$, while the DNS shows $\Reyy{} = \Or{(\Ra{}})$. The weak flow can be shown to be dominated by the $n=2$ mode of the velocity field in \cref{uvFourier}, not captured by the RD model since it truncates velocity at $n = 0$. A more detailed analysis of this conductive state will be included in our future work. 

The most important question now becomes whether the new RD model accurately predicts thermal transport as this was the shortcoming of the ODE model of \citep{Moore_Huang_2024}. Accordingly, \Cref{fig6}(c) shows measurements of the Nusselt number from the DNS (data points), the new RD model (solid curve), and the scaling law $\Nu{} \propto \Ra{}^{1/4}$ obtained from boundary layer analysis. The figure confirms that the RD model captures the $\Nu$ computed in the full DNS for all three Prandtl numbers tested. The RD model shows that $\Nu{}$ is unity for $\Ra{}<\Ra_1^*$ and then rises sharply for $\Ra{}>\Ra_1^*$ as thermal convection sets in and strongly enhances thermal transport. The discrepancies between the RD model and DNS are relatively small, on the order of 5-20\%, over 4 decades of $\Ra{}>\Ra{}_1^*$ and 1.5 decades of $\Pra{}$. This level of accuracy is perhaps better than expected considering that the RD model only resolves the lowest non-trivial modes, $u_0(r,t), a_1(r,t), b_1(r,t)$, in the $\theta$ variable. 
Furthermore, \cref{fig6}(c) shows that, for large values of $\Ra{}$, both the RD model and the full DNS are consistent with the $\Nu{} \propto \Ra{}^{1/4}$ power law, \cref{Nu-scaling}, that arises from boundary layer analysis. This agreement suggests that convective thermal transport in the annulus primarily arises from the boundary-layer structure of the lowest Fourier modes, $u_0(r,t), a_1(r,t), b_1(r,t)$, near the thermally-driven boundary, $r=1/2$. Thermal transport can be accurately predicted by the inner-outer matching problem formulated in the previous section, a key improvement over the ODE model \citep{Moore_Huang_2024} which does not resolve the boundary-layer and incorrectly predicts $\Nu{}$ to saturate to a constant at high $\Ra{}$.

\begin{figure}
 \includegraphics[width = 0.8\textwidth]{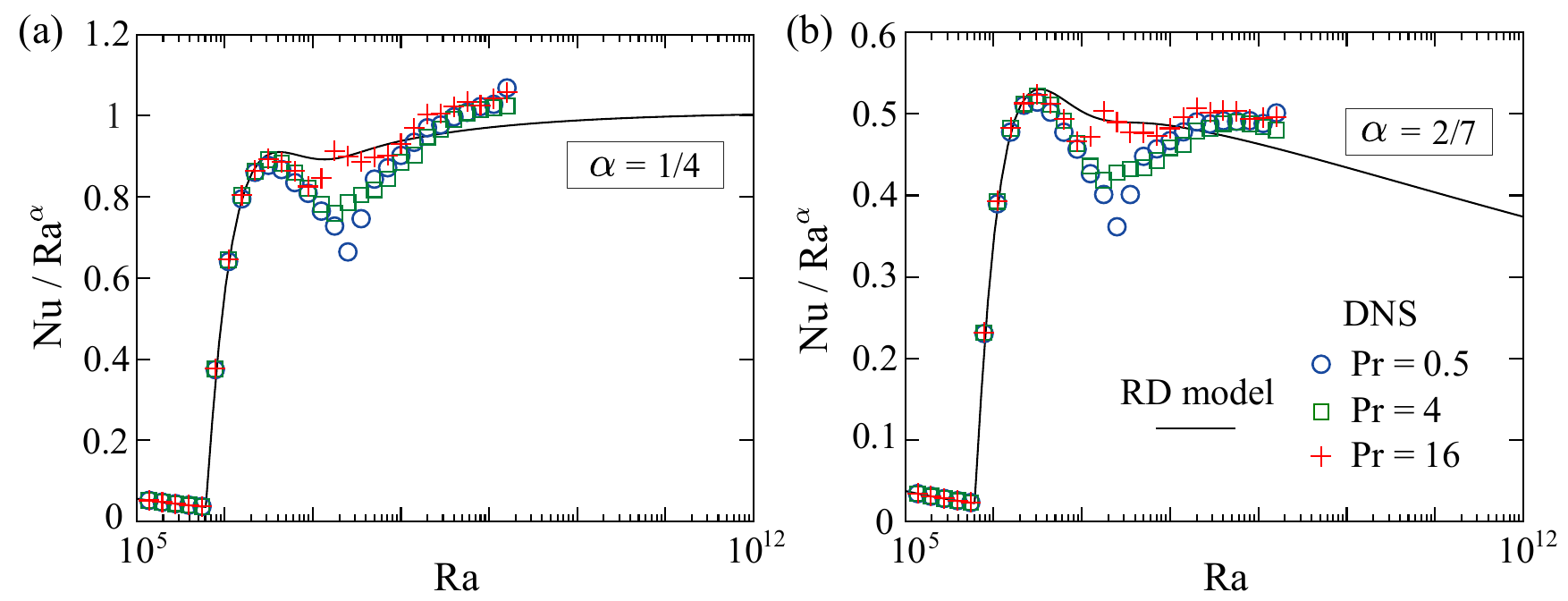}
 \centering
 \caption{Compensated plot of $\Nu{}/\Ra{}^\alpha$, where the exponent $\alpha$ is taken as $1/4$ in (a) and $2/7$ in (b). The DNS data and RD model solution in (a)-(b) are identical to those shown in \cref{fig6}(c).
 }
\label{fig7}
\end{figure}

The agreement in \cref{fig6}(c) suggests that these lowest $\theta$-modes are the ones most responsible for thermal transport. Higher-order modes and dynamic features such as LSC reversals may bring further enhancement to the thermal transport and may potentially alter the scaling of $\Nu{}$, for example producing a $2/7$ law \citep{Castaing_1989}. To examine this possibility, we show in \cref{fig7} compensated plots of $\Nu{}/\Ra{}^\alpha$ , where $\alpha$ is chosen as $1/4$ and $2/7$ in \cref{fig7}(a) and (b) respectively. The RD model shows $\alpha\to 1/4$ for large $\Ra{}$ as the compensated data approach a horizontal asymptote in \cref{fig7}(a), consistent with the boundary layer analysis. The DNS data also appear broadly consistent with $\alpha=1/4$, although the $\Nu{}/\Ra{}^{1/4}$ measurements move above the RD data for the largest $\Ra{}$ run. It is unclear whether this compensated data is leveling off at high $\Ra{}$ or if it is continuing to grow slightly. To examine an alternative, \cref{fig7}(b) shows compensated data with $\alpha = 2/7$. The figure suggests that these compensated data points come  closer to leveling off overall, although the data for $\Pr{}=4$ and $\Pr{}=16$ do show a slight downward trend at the highest $\Ra{}$ (around $10^9$).
In summary, it is difficult to distinguish between the two similar power laws $\alpha=1/4$ and $\alpha=2/7$, and the DNS data is broadly consistent with either law. One may be tempted to perform a least-squares  procedure to infer a specific best-fit $\alpha$, but such a procedure would necessarily involve choices, such as the range of $\Ra{}$ to be fit, and hence would be not be as objective as imagined. Regardless, the RD model, being a substantially reduced model, surely misses many features of convection that contribute to thermal transport. The 5-20\% discrepancies in $\Nu{}$ we attribute primarily to the higher-order Fourier modes and the dynamical LSC reversals. These factors contribute positively to the $\Nu{}$ and are responsible for features such as bulk mixing and unsteady boundary layer effects that are not included in the low-order RD model.

\section{Discussions}
\label{discussion}

In this manuscript we have extended the ODE model developed by Moore \& Huang \citep{Moore_Huang_2024} for convection in an annulus to a new PDE model that fully resolves radial dependence of velocity and temperature fields while only retaining the lowest Fourier modes in the angular direction. The new model takes the form of a reaction-diffusion system. While this new model fails to predict dynamic events, such as LSC reversals, it substantial improves the prediction of the time-averaged thermal transport. In particular, boundary-layer analysis of the new model predicts the scaling relationship $\Nu{} \propto \Ra{}^{1/4}$, which is shown to be in good agreement with direct numerical simulations, although it is difficult to distinguish from the scaling $\Nu{} \propto \Ra{}^{2/7}$.

We note that the $\Nu{}$--$\Ra{}$ scaling depends on factors such as the domain geometry and the values of $\Ra{}$ and $\Pra{}$. Thus, a simple power law with a rational exponent cannot be expected to hold in general \cite{grossmann_lohse_2013}. From the Grossmann-Lohse theory for the convection in a rectangular domain \citep{grossmann2000scaling, RevModPhys.81.503, grossmann_lohse_2013}, a power law of the form $\Nu{}\propto \Ra{}^\alpha$ may hold locally, but the exponent $\alpha$ depends on $\Pra{}$ and the range of $\Ra{}$ under consideration. We remark that important differences exists between the rectangular and annular domains, namely  the inner radius of the annulus suppresses bulk motion through the center of the domain. An advantage of this domain is that it enables one to examine the interplay of large-scale circulation and thermal boundary layers in isolation from other effects, such as bulk motion through the center or corner rolls. In this geometry, we have shown that it is possible to develop a simple, first-principled theory yielding a scaling law $\Nu{} \propto \Ra{}^{1/4}$ that is in good agreement with DNS. This theory may offer a baseline understanding of thermal transport since it only depends on the bulk circulatory flow and boundary-layer effects near the thermal driven surface. 
This approach of systematically analyzing PDEs that arise directly from the governing equations may complement other more elaborate theories that rely on phenomenology or dimensional analysis to fill in unknowns. Furthermore, it may be possible to extend the current theory in a systematic manner by including additional Fourier modes. The inclusion of higher modes might even account for turbulent fluctuations without the need to make additional postulates about the nature of those fluctuations.

A shortcoming of the new RD model is that by over-resolving the radial dependence compared to the angular dependence, the damping action of diffusion suppresses the LSC reversal events that were accurately captured by the cruder ODE model. This shortcoming may perhaps be overcome by including additional $\theta$-modes in model, thereby increasing the number of coupled PDEs. How many additional modes must be included to recover the reversals is a question for near-future research. Alternatively, it might be possible to account for the higher-order modes through stochastic forcing, with the forcing term chosen judiciously to recover statistical features of the system.
With these questions left for future research, the key finding of the current paper is that the resolution in the radial direction does significantly improve time-averaged thermal transport.
This improvement, for example, permits the model to be coupled to interfaces that move in response to thermal or diffusive transport, such as objects melting or dissolving in convectively excited flows \cite{Huang2015, Moore2017, Weady2022, Huang2022, zukowski2025invariant}.

\section*{\si{}}
Supplementary movies are available at \url{https://math.nyu.edu/~jinzi/research/AnnularConvectionHeatTransfer/}.
\bibliography{manuscript}

\end{document}